\documentclass[aip,
nofootinbib,
rsi,%
reprint,
twocolumn,
longbibliography,
author-numerical%
]{revtex4-2}

\usepackage[utf8]{inputenc}
\usepackage{amsfonts,amssymb,amsmath}
\usepackage[usenames]{color}
\usepackage{graphicx}
\usepackage{changes,comment}

\usepackage{natbib}

\newcommand{\pdr}[2]{\dfrac{\partial{#1}}{\partial {#2}}}
\newcommand{\pddr}[2]{\dfrac{\partial^2{#1}}{\partial {#2}^2}}

\newcommand{\pdra}[2]{{\partial{#1}}/{\partial{#2}}}
\newcommand{\pddra}[2]{{\partial^2{#1}}/{\partial{#2}^2}}

\newcommand{\tx}{\tilde{x}}

\newcommand{\tZ}{\widetilde{Z}}
\newcommand{\tT}{\widetilde{T}}
\newcommand{\tR}{\widetilde{R}}

\newcommand{\tj}{\tilde{j}}
\newcommand{\teta}{\tilde{\eta}}

\newcommand{\Esig}{T_{\sigma}}

\newcommand{\tc}{\tilde{c}}
\newcommand{\cref}{c_{ref}}
\newcommand{\Cdl}{C_{dl}}

\newcommand{\veps}{\varepsilon}

\newcommand{\lexp}[1]{\exp\left(#1\right)}

\newcommand{\sion}{\sigma_p}

\newcommand{\lcat}{l_t}
\newcommand{\tit}{\tilde{t}}
\newcommand{\tom}{\tilde{\omega}}
\newcommand{\ri}{{\rm i}}
\newcommand{\expo}{{\rm e}^{\teta_0}}

\newcommand{\etal}{{ }et al.{ }}

\begin{document}

\sf

\title{The effect of in-phase current and temperature oscillations on the impedance
      of the cathode catalyst layer in a PEM fuel cell}

\author{Andrei Kulikovsky}
\email{A.Kulikovsky@fz-juelich.de}

\affiliation{Forschungszentrum J\"ulich GmbH           \\
    Theory and Computation of Energy Materials (IET--3)   \\
    Institute of Energy and Climate Research,              \\
    D--52425 J\"ulich, Germany
}

\date{\today}

\begin{abstract}
An impedance model for the cathode catalyst layer (CCL) in a PEM fuel cell
demonstrates that in-phase harmonic perturbations to the cell current density
and CCL temperature lower both CCL impedance and static resistivity.
This mitigation is primarily driven by the oscillating exchange current
density of the oxygen reduction reaction.
\end{abstract}


\begin{center}

\end{center}

\maketitle

\section{Introduction}

Electrochemical impedance spectroscopy (EIS) is one of the best and widely used
tools for polymer electrolyte membrane (PEM) fuel cells  characterization
(see Lasia's book~\cite{Lasia_book_14} and reviews~\cite{Zhang_20,Huang_20}).
A classic EIS is based on application of a small harmonic perturbation to the cell
current density and measuring the response of the cell potential.
A promising and not well studied option is the EIS-like regime of cell
operation, with oscillations applied to one or two of the cell operating parameters.
Experiments of Kim \etal\cite{Kim_08b} and Hwang \etal\cite{Hwang_10}
have shown improvement of the PEMFC polarization curve
when the cathode flow velocity was oscillating.
The model developed by Kulikovsky\cite{Kulikovsky_24a,Kulikovsky_24d} has
demonstrated a dramatic reduction in PEMFC resistivity when
in-phase harmonic perturbations of the cell potential and oxygen concentration
were applied. Recently, a model of Kulikovsky\cite{Kulikovsky_26d} has shown
a reduction of the the cathode catalyst layer (CCL) impedance
due to the lowering of proton transport losses upon application of the
in-phase cell current density and temperature perturbations.

Below, we demonstrate that in-phase oscillations of the cell current density
and temperature decrease the impedance of the CCL much more than
was shown in\cite{Kulikovsky_26d}.  The temperature oscillations induce
the oscillations in the CCL proton conductivity $\sion$ and the exchange
current density $i_*$ of the oxygen reduction reaction (ORR). The
oscillating $\sion$ lowers the proton transport losses, while
the oscillating $i_*$ reduces the faradaic impedance.
The dominating contribution to the impedance reduction gives the oscillating $i_*$.
In the limit of zero frequency of the applied perturbations,
the static cell resistivity is also reduced.

\section{Model}

Consider the cathode catalyst layer of a thickness $\lcat$. Let the $x$-axis
with the origin at the membrane interface be directed toward the gas diffusion layer (GDL).
To simplify the analytical model, we will neglect the oxygen transport losses in the CCL.

The proton charge conservation equation is
\begin{equation}
   \Cdl\pdr{\eta}{t} - \sion\pddr{\eta}{x}
      = - i_*\left(\dfrac{c_1}{\cref}\right)\lexp{\dfrac{\eta}{b}}
   \label{eq:etax}
\end{equation}
Here,
$\Cdl$ is the double layer capacitance,
$\eta$ is the positive by convention ORR overpotential,
$\sion$ is the CCL proton conductivity,
$i_*$ is the ORR volumetric exchange current density,
$c_1$ is the uniform oxygen molar concentration in the CCL,
$\cref$ the reference concentration, and
$b$ is the ORR Tafel slope.

Two parameters in Eq.\eqref{eq:etax} exponentially change with the temperature:
the CCL proton conductivity $\sion$ and the ORR exchange current density $i_*$.
Springer, Zawodzinski and Gottesfeld\cite{Springer_91} reported the Arrhenius law
for the temperature dependence $\sion(T)$. For the exchange current density
$i_*(T)$ of the ORR running in the Pt/C-based catalyst
layer with Nafion as a proton conductor, the Arrhenius law has been measured by
Parthasarathy, Srinivasan and Appleby\cite{Srini_92}:
\begin{equation}
   \begin{split}
    & \sigma_p(T) = \sion^*\exp\left(\Esig\left(\dfrac{1}{303} - \dfrac{1}{T}\right) \right),
      \quad \Esig = 1268~\text{K.} \\
    & i_*(T) = i_{*,0}\exp\left(T_*\left(\dfrac{1}{353} - \dfrac{1}{T}\right) \right),
           \quad T_* = 8807~\text{K}.
   \end{split}
   \label{eq:sionT}
\end{equation}
Below, the superscripts 0 and 1 mark the steady-state values and the
small perturbation amplitudes, respectively.
Setting $\sion = \sion^0 + \sion^1$, $i_* = i_*^0 + i_*^1$,
$T = T^0 + T^1$, expanding the exponents in a Taylor series
and retaining the two leading terms, we get the linear relations
between the temperature $T^1$ and  $\sion^1$, $i_*^1$ perturbations:
\begin{equation}
      \dfrac{\sion^1}{\sion^0} = \dfrac{\Esig T^1}{\left(T^0\right)^2}, \quad
      \dfrac{i_*^1}{i_*^0} = \dfrac{T_* T^1}{\left(T^0\right)^2}
   \label{eq:sion1}
\end{equation}
where  $\sion^0 \equiv \sion(T^0)$, $i_*^0 \equiv i_*(T^0)$,
and $T^0$ is the steady-state cell temperature, which is assumed to
be constant through the CCL.
Due to linearity, Eqs.\eqref{eq:sion1} hold also
for the perturbation amplitudes $\sion^1(\omega)$, $i_*^1(\omega)$
and $T^1(\omega)$ in the frequency domain.

Setting in Eq.\eqref{eq:etax} $\eta = \eta^0(x) +\eta^1(x,t)$, $\sion = \sion^0 + \sion^1(t)$,
$i_* = i_*^0 + i_*^1(t)$,
expanding exponent in a Taylor series, neglecting the term with the perturbation products,
and subtracting the static equation for $\eta^0$
\begin{equation}
   \sion^0\pddr{\eta^0}{x} =  i_*^0\left(\dfrac{c_1}{\cref}\right)\lexp{\dfrac{\eta^0}{b}}
   \label{eq:eta0x}
\end{equation}
we get a linear equation for the small perturbation amplitude $\teta^1$ in the time domain
\begin{multline}
   \Cdl\pdr{\eta^1}{t} - \sion^1\pddr{\eta^0}{x} - \sion^0\pddr{\eta^1}{x} \\
       = - i_*^0\left(\dfrac{c_1}{\cref}\right)\lexp{\dfrac{\eta^0}{b}}\dfrac{\eta^1}{b}
         - i_*^1\left(\dfrac{c_1}{\cref}\right)\lexp{\dfrac{\eta^0}{b}}
   \label{eq:eta1x}
\end{multline}
Using the Fourier-transforms $\eta^1(x,t) = \eta^1(x, \omega)\exp(\ri\omega t)$,
$\sion^1(t) = \sion^1(\omega)\exp(\ri\omega t)$, $i_*^1(t) = i_*^1(\omega)\exp(\ri\omega t)$
from Eq.\eqref{eq:eta1x} we obtain
the equation for the perturbation amplitude $\eta^1(x,\omega)$ in the frequency domain
\begin{multline}
   \Cdl\ri\omega\eta^1 - \sion^1\pddr{\eta^0}{x} - \sion^0\pddr{\eta^1}{x} \\
       = - i_*^0\left(\dfrac{c_1}{\cref}\right)\lexp{\dfrac{\eta^0}{b}}\dfrac{\eta^1}{b}
        - i_*^1\left(\dfrac{c_1}{\cref}\right)\lexp{\dfrac{\eta^0}{b}}
   \label{eq:eta1xF}
\end{multline}
Using Eq.\eqref{eq:eta0x} to eliminate $\pddra{\eta^0}{x}$,
rearranging the terms, and taking into account  Eqs.\eqref{eq:sion1},
from Eq.\eqref{eq:eta1xF} we find
\begin{multline}
   \sion^0\pddr{\eta^1}{x}  =  \Cdl\ri\omega\eta^1 \\
       + \left(\dfrac{\eta^1}{b} + \dfrac{T_* T^1}{\left(T^0\right)^2}
                                 - \dfrac{\Esig T^1}{\left(T^0\right)^2}\right)
                                    i_*^0\left(\dfrac{c_1}{\cref}\right)\lexp{\dfrac{\eta^0}{b}}
   \label{eq:eta1xF3}
\end{multline}

The boundary conditions for Eq.\eqref{eq:eta1xF3} are
\begin{equation}
   \left(- \sion^0 \pdr{\eta^1}{x} - \sion^1 \pdr{\eta^0}{x}\right)_{x=0}  = j^1, \quad
   \left.\pdr{\eta^1}{x}\right|_{x=\lcat} = 0
   \label{eq:bc1}
\end{equation}
where the first equation follows from linearization of the Ohm's law
$- (\sion^0 +\sion^1)\pdra{(\teta^0 + \teta^1)}{\tx} = j_0 + j_1$.
Using the static Ohm's law $- \sion^0\pdra{\eta^0}{x}|_{\tx=0} = j_0$
and Eq.\eqref{eq:sion1}, Eq.\eqref{eq:bc1} is transformed to
\begin{equation}
   - \left.\sion^0 \pdr{\eta^1}{x} \right|_{x=0}
       = j^1 - \dfrac{\Esig T^1}{\left(T^0\right)^2}\,j_0, \quad
   \left.\pdr{\eta^1}{x}\right|_{x=\lcat} = 0
   \label{eq:bc3}
\end{equation}

To simplify calculations, we introduce the dimensionless variables
\begin{multline}
   \tx = \dfrac{x}{\lcat}, \quad \tit = \dfrac{t}{t_*}, \quad \teta = \dfrac{\eta}{b},
   \quad \tj = \dfrac{j}{j_*}, \quad \tT = \dfrac{T}{\Esig}, \\
   \tom = \omega t_*, \quad \tZ = \dfrac{Z \sion}{\lcat}
   \label{eq:dless}
\end{multline}
where $\omega$ is the AC angular frequency, $Z$ the impedance, and
\begin{equation}
    t_* = \dfrac{\Cdl b}{i_*^0}, \quad j_* = \dfrac{\sion^0 b}{\lcat}.
    \label{eq:tast}
\end{equation}
are the characteristic time and current density, respectively.
With Eqs.\eqref{eq:dless}, Eqs.\eqref{eq:eta1xF3}, \eqref{eq:bc1} take the form
\begin{equation}
   \veps^2\pddr{\teta^1}{\tx}  =  \left(\ri\tom +  \tc_1\expo\right)\teta^1
       - (1 - \beta_T)\Lambda^1\tc_1\expo
   \label{eq:teta1xF3}
\end{equation}
\begin{equation}
   - \left. \pdr{\teta^1}{\tx} \right|_{\tx=0}  = \tj^1 - \Lambda^1\tj_0, \quad
   \left.\pdr{\teta^1}{\tx}\right|_{\tx=1} = 0
   \label{eq:tbc3}
\end{equation}
where
\begin{equation}
   \Lambda^1 = \tT^1 \bigg/ \left(\tT^0\right)^2,
   \label{eq:Lam1}
\end{equation}
and $\veps$, $\beta_T$ are the constant parameters:
\begin{equation}
   \veps = \sqrt{\dfrac{\sion^0 b}{i_*^0 \lcat^2}}, \quad \beta_T = \dfrac{T_*}{\Esig} \simeq 6.95.
   \label{eq:veps}
\end{equation}

\section{Results and discussion}

Let the cell current density be small. In that case, we can neglect the variation of the ORR
overpotential through the CCL depth and set $\teta^0 = \teta_0$, where
$\teta_0$ is the overpotential at the membrane surface ($\tx=0$).
The subscripts 0 and 1 denote the membrane/CCL and CCL/GDL interface, respectively.
The dimensionless Tafel law
\begin{equation}
   \veps^2 \tj_0 = \tc_1\expo
   \label{eq:tTafel}
\end{equation}
allows us to replace  $\tc_1\expo$ in Eq.\eqref{eq:teta1xF3} with $\veps^2 \tj_0$.
The solution to Eq.\eqref{eq:teta1xF3} with the boundary conditions Eq.\eqref{eq:tbc3} is
\begin{multline}
   \teta^1(\tx) = \dfrac{\tj^0\Lambda^1 - \tj^1}{\phi}\bigl(\sinh(\phi\tx) - \coth(\phi) \cosh(\phi\tx)\bigr) \\
               + \dfrac{\tj_0 (1 - \beta_T) \Lambda^1}{\phi^2},
               \quad \phi = \sqrt{\tj_0 + \ri\tom/\veps^2}
   \label{eq:teta1_sol}
\end{multline}

From Eq.\eqref{eq:teta1_sol} we get the CCL impedance $\tZ = \teta^1/\tj^1 |_{\tx=0}$:
\begin{equation}
   \tZ = (1 - \kappa)\dfrac{\coth\sqrt{\tj_0 + \ri\tom/\veps^2}}{\sqrt{\tj_0 + \ri\tom/\veps^2}}
   + \dfrac{\kappa (1 - \beta_T)}{\tj_0 + \ri\tom/\veps^2}
   \label{eq:tZtot}
\end{equation}
where $\kappa$ is the temperature control parameter:
\begin{equation}
   \kappa =  \dfrac{\Lambda^1\tj_0}{\tj^1}
   \label{eq:kap}
\end{equation}
To estimate a realistic value of $\kappa$, we take $T^1 = 0.1$~K, $j_0 = 1$~A~cm$^{2}$,
$j^1 = 0.01$~A~cm$^{-2}$ and $T^0 = 353$~K. With these parameters and
$\Esig = 1268$~K, we get $\kappa \simeq 0.1$.
Thus, by selecting appropriate values of the amplitudes  $\tj^1$ and $\Lambda^1$,
the control parameter $\kappa$ can be set to a value
in the range $\kappa \in [0, 1]$, which is considered below.

\begin{figure}
\begin{center}
   \includegraphics[scale=0.45]{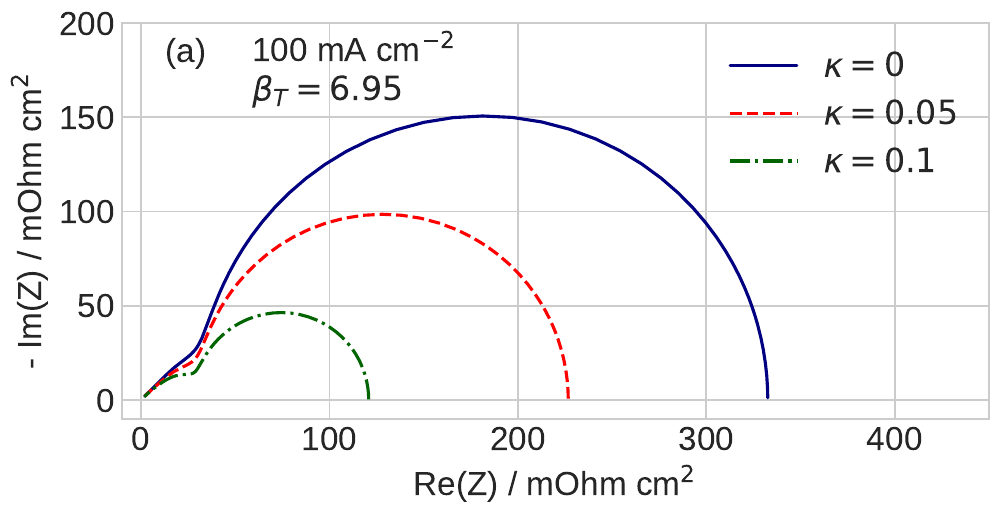}\\
   \includegraphics[scale=0.45]{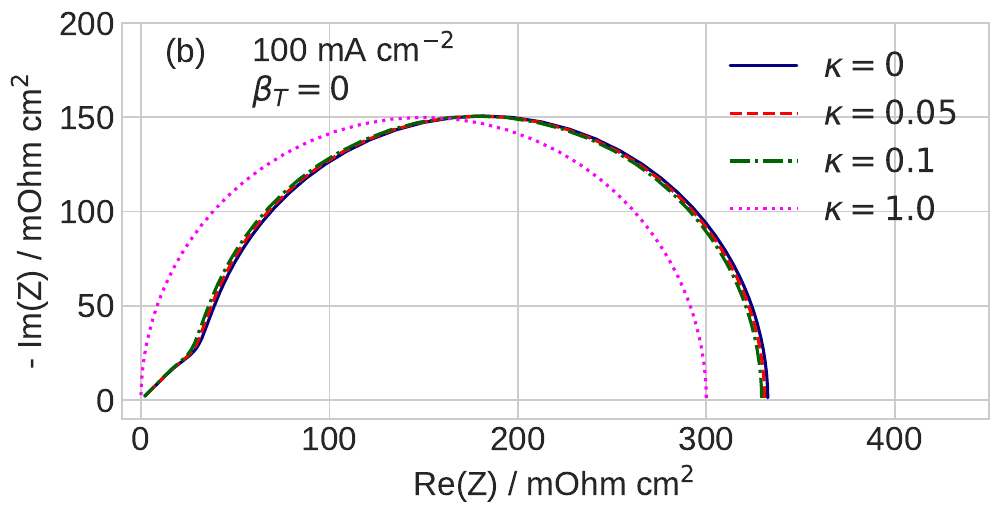}
\end{center}
\caption{(a) The Nyquist spectra of impedance Eq.\eqref{eq:tZtot} for the three values
     of the temperature control parameter $\kappa$. The cell current density
     is 100 mA~cm$^{-2}$. The oscillations of the ORR exchange current density
     are ``switched on'', $\beta_T=6.95$.
     (b) The Nyquist plots of impedance Eq.\eqref{eq:tZtot} with the
     oscillations of the exchange current ``switched off'', $\beta_T=0$.
}
\label{fig:Nyq100}
\end{figure}

At zero temperature perturbation, $\Lambda^1=0$, $\kappa=0$,
and Eq.\eqref{eq:tZtot} reduces to the Warburg-like impedance of the
CCL with the finite rate of proton transport\cite{KulikovskY_13a}:
\begin{equation}
   \tZ_{fp} = \dfrac{\coth\sqrt{\tj_0 + \ri\tom/\veps^2}}{\sqrt{\tj_0 + \ri\tom/\veps^2}}
   \label{eq:tZccl}
\end{equation}
The blue solid curve in Figure~\ref{fig:Nyq100}a shows
the typical shape of the Nyquist spectrum of $Z_{fp}$
for the set of parameters in Table~\ref{tab:parms}.
The spectrum comprises the faradaic arc connected to the high-frequency $45^\circ$ straight
line, which exhibits proton transport impedance.

\begin{table}
\small
\begin{tabular}{|l|c|}
\hline
     ORR exchange current density $i_*$,  A~cm$^{-3}$ &  $ 10^{-4}$ \\
     Double layer capacitance $\Cdl$, F~cm$^{-3}$ &  20 \\
     ORR Tafel slope $b$, mV/exp               & 30 \\
     CCL proton conductivity $\sion^0$, $\Omega^{-1}$~cm$^{-1}$  & 0.01 \\
     CCL thickness $\lcat$, $\mu$m           & 10   \\
\hline
   	 Cathode pressure $p$, bar                 & 1.0 \\
     Cell temperature $T^0$, K                 & $273 + 80$ \\
     Cell current density $j_0$, A~cm$^{-2}$   &  0.1, 0.413  \\

\hline
\end{tabular}
\caption{The cell properties and operation parameters.}
\label{tab:parms}
\end{table}

Now consider the case of non-zero temperature perturbation, $\kappa > 0$.
For a moment, we leave aside the question of how the high-frequency
temperature perturbation could be produced. The Nyquist plots of
the impedance $Z$, Eq.\eqref{eq:tZtot}, for $\kappa$ of 0.05 and 0.1
and the parameters in Table~\ref{tab:parms} are shown in Figure~\ref{fig:Nyq100}a.
As can be seen, a positive kappa reduces both the size of the faradaic arc and
that of the straight proton transport line.
The most striking feature of the curves in Figure~\ref{fig:Nyq100}a is that
the {\em static} cell resistivity (the rightmost
point of the Nyquist spectrum) decreases with $\kappa$.

It is advisable to separate the effects of the conductivity and exchange
current density perturbations. By setting $\beta_T=0$ we ``switch off'' the ORR exchange
current density oscillations. With  $\kappa = 1$,
Eq.\eqref{eq:tZtot} reduces to\cite{Kulikovsky_26d}
\begin{equation}
   \tZ_{RC} = \dfrac{1}{\tj_0 + \ri\tom/\veps^2},
   \label{eq:tZrc}
\end{equation}
which is an ideal semicircle representing the parallel $RC$-circuit impedance
(the dotted magenta line in Figure~\ref{fig:Nyq100}b).
In dimensional form, Eq.\eqref{eq:tZrc} does not contain proton conductivity,
i.e., at $\kappa=1$, the in-phase temperature and
current density oscillations fully eliminate proton transport losses in the CCL.

This effect can be explained as follows.
A real and positive $\kappa$ means that $T^1$ is in phase with $j^1$.
According to Eq.\eqref{eq:sion1}, the resulting proton conductivity oscillations
are also in phase with the $j^1$ oscillations. Writing the Ohm's law in the form
\begin{equation}
   - \sion^0 \pdr{\eta^1}{x} = j^1 + \sion^1 \pdr{\eta^0}{x}
                             = j^1 - \dfrac{\sion^1}{\sion^0} j_0
   \label{eq:Ohms}
\end{equation}
we see that the conductivity oscillations $\sion^1$ reduce
the amplitude of $\pdra{\eta^1}{x}$ oscillations. Furthermore,
when $\sion^1 j_0/\sion^0 = j^1$, which is equivalent to $\kappa=1$,
the right side of Eq.\eqref{eq:Ohms} is zero and we get $\pdra{\eta^1}{x} = 0$.
A uniform along the $\tx$-axis overpotential perturbation $\teta^1$  means
that there are no proton transport losses in the system.

Figure~\ref{fig:Nyq100}b also shows the Nyquist spectra Eq.\eqref{eq:tZtot}
with $\beta_T=0$ and the same values of $\kappa$ as in Figure~\ref{fig:Nyq100}a.
As can be seen, at small $\kappa$ the reduction of impedance due to
proton conductivity oscillations is rather marginal.

A comparison of Figures~\ref{fig:Nyq100}a and b
shows that the main effect of the CCL impedance reduction
is due to the oscillating  exchange current density $i_*$.
The Arrhenius slope of $i_*$ variation with temperature is nearly seven times
larger than the proton conductivity slope, which explains the dominant
effect of $i_*$ perturbations. Pumping the CCL temperature in phase
with the cell current dramatically improves the ORR rate and reduces
the faradaic cell resistivity.

Setting in Eq.\eqref{eq:tZtot} $\tom =0$ we find the CCL static resistivity
\begin{equation}
   \tR = (1 - \kappa) \dfrac{\coth\sqrt{\tj_0}}{\sqrt{\tj_0}}
      + \dfrac{\kappa (1 - \beta_T)}{\tj_0}
   \label{eq:tR}
\end{equation}
At small $\tj_0$, $\coth\sqrt{\tj_0} \simeq 1 / \sqrt{\tj_0} + \sqrt{\tj_0}/3$, and we get
\begin{equation}
    \tR = \dfrac{1}{\tj_0} + \dfrac{1}{3}
       - \kappa\dfrac{\beta_T}{\tj_0} - \kappa\dfrac{1}{3}
    \label{eq:tRsmall}
\end{equation}
which in dimensional form is
\begin{equation}
    R = \dfrac{b}{j_0} + \dfrac{\lcat}{3\sion}
       - \kappa\beta_T \dfrac{b}{j_0} - \kappa\dfrac{\lcat}{3\sion}
    \label{eq:Rsmall}
\end{equation}
Here, the first term on the right side is the faradaic resistivity and
the second term is the proton transport resistivity. The third and
fourth terms describe the reduction in faradaic and proton transport
resistivities, respectively, due to temperature oscillations.
Reducing $\omega$ while keeping $\kappa$ constant transfers the CCL
to the steady state, where the static resistivity is given by Eq.\eqref{eq:Rsmall}.

At higher cell currents, the effect can be demonstrated using Eq.\eqref{eq:teta1xF3}
with the $\tx$-dependent static overpotential $\teta^0(\tx)$. The latter
is obtained by solving Eq.\eqref{eq:eta0x}, which, in dimensionless form, reads
\begin{equation}
   \veps^2\pddr{\teta^0}{\tx} =  \tc_1\exp\teta^0,
      \quad \teta^0(0) = \teta_0, \quad \left.\pdr{\teta^0}{\tx}\right|_{\tx=0} = 0
   \label{eq:teta0x}
\end{equation}
The solution of the BVP problem Eq.\eqref{eq:teta0x}
should be substituted into Eq.\eqref{eq:teta1xF3}. Solving Eq.\eqref{eq:teta1xF3} numerically
we get the system impedance $\tZ = \teta^1/\tj^1 |_{\tx=0}$ at higher currents.

\begin{figure}
\begin{center}
   \includegraphics[scale=0.45]{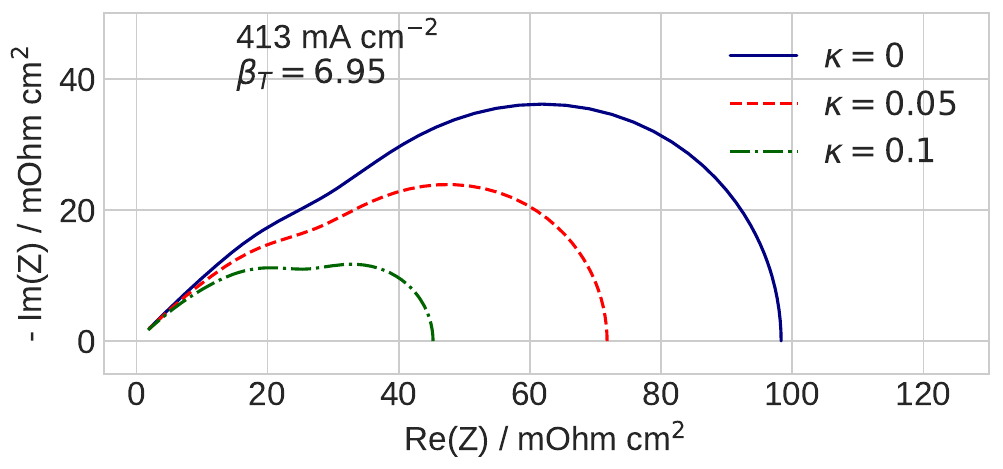}
\end{center}
\caption{The Nyquist spectra of the CCL impedance resulting from the numerical
solution of Eq.\eqref{eq:teta1xF3}
     with the nonuniform along the $\tx$--coordinate static overpotential $\teta^0(\tx)$
     for the three values
     of the temperature control parameter $\kappa$. The cell current density
     is 413 mA~cm$^{-2}$.
}
\label{fig:Nyq413}
\end{figure}

Figure~\ref{fig:Nyq413} shows the Nyquist spectra obtained in this way
for a cell current density of 0.413 A~cm$^{-2}$.
Without temperature control, the spectrum resembles a Warburg finite-length
impedance (the solid blue curve in Figure~\ref{fig:Nyq413}).
With positive $\kappa$, both the high-frequency proton transport
line, and the main faradaic arc decrease in size (Figure~\ref{fig:Nyq413}).

The model above ignores the finite relaxation times of $\sion$ and $i_*$
upon a change in the CCL temperature. However, at sufficiently small frequencies
of the applied perturbation, the relaxation times are much shorter than
the AC signal period, and Eqs.\eqref{eq:sionT} can be used. While the finite
relaxation times of $\sion$ and $i_*$ could alter the shape of
the Nyquist spectrum, the static limit would still be given by Eq.\eqref{eq:Rsmall}.

Low-frequency CCL temperature oscillations can be created by attaching the heating
pad to the external surface of the cathode flow field.  The temperature controller
of the heating pad should be set to eliminate the phase shift
between the cell current and CCL temperature oscillations.

\newpage

\end{document}